# Diffraction of Gaussian beam in a 3D smoothly inhomogeneous media: eikonal-based complex geometrical optics approach

P. Berczynski[1], K.Yu. Bliokh[2,3], Yu.A. Kravtsov[4,5], and A. Stateczny[5]

[1]*Institute of Physics, Szczecin University of Technology, Piastow 19, 70310 Szczecin, Poland*
[2]*Institute of Radio Astronomy, 4 Krasnoznamyonnaya st., 61002 Kharkov, Ukraine*
[3]*A.Ya. Usikov Institute of Radiophysics and Electronics, 12 Akademika Proskury st., 61085 Kharkov, Ukraine*
[4]*Space Research Institute, Russian Academy of Sci., Profsoyuznaya st. 82/34, 117997 Moscow, Russia*
[5]*Institute of Physics, Maritime University of Szczecin, Waly Chrobrego, 70500 Szczecin, Poland*

The paper presents an ab initio account of the paraxial complex geometrical optics (CGO) in application to a scalar Gaussian beam propagation and diffraction in a 3D smoothly inhomogeneous medium. The paraxial CGO deals with quadratic expansion of the complex eikonal and reduces the wave problem to the solution of ordinary differential equations of Riccati type. This substantially simplifies description of Gaussian beams diffraction as compared to full wave or parabolic (quasi-optics) equations. For a Gaussian beam propagating in a homogeneous medium or along the symmetry axis in a lens-like medium, the CGO equations possess analytical solutions, otherwise they can be readily solved numerically. As a non-trivial example we consider Gaussian beam propagation and diffraction along a helical ray in an axially symmetric waveguide medium. It is shown that the major axis of the beam's elliptical cross-section grows unboundedly; it is oriented predominantly in azimuthal (binormal) direction, and does not obey the parallel transport law.

## 1. Introduction

Conventional real geometrical optics is a method designed to describe trajectories of the rays, along which the phase and amplitude of a wave field can be calculated in the diffractionless approximation [1,2]. Complex extension of the geometrical optics theory enables one to include diffraction processes into the scope of consideration, which characterize the wave rather than the geometrical features of the wave beams (by diffraction we here mean diffraction spreading of the wave beam). There are two main forms of the complex geometrical optics (CGO): the ray-based form, which deals with complex rays, i.e. trajectories in complex space, and the eikonal-based form, which uses complex eikonal instead of the complex rays [2–4]. The ability of CGO to describe diffraction of Gaussian beam (GB) on the basis of complex Hamiltonian ray equations was already demonstrated 30–35 years ago within the framework of the ray-based approach. According to [5–7] (see also [2,3]), CGO ensures analytical results equivalent to the exact solution of parabolic wave equation. Development of numerical methods within the framework of the ray-based CGO in recent years allowed to describe GB propagation and diffraction in inhomogeneous media: GB focusing by localized inhomogeneities [8,9], reflection from a linear-profile layer [10] as well as other issues.

The eikonal-based form of the paraxial CGO seems to be an even more powerful instrument of the wave theory as compared with the ray-based version. It reduces GB diffraction problem to the solution of ordinary differential equations of Riccati type, which are more convenient for numerical solution than Hamiltonian ray equations.

Development of the eikonal-based form of CGO was essentially influenced by *Babič approach* [11–13], which deals with narrow beams, concentrated in the vicinity of the central ray (which is the geodesic in the space with the corresponding metric). Babič approach is based on an abridged parabolic wave equation, which preserves only quadratic terms in small deviations from a central ray. There the solution of the abridged parabolic wave equation has been obtained in the form of GB with parameters obeying complex-valued Riccati equation. Thereby Babič reduced GB diffraction problem to solution of ordinary differential equations of Riccati type. Babič approach gave rise to the *dynamical ray tracing* method, which also reduces diffraction problems to solution of Riccati-type equations on the basis of different techniques [14–16]. Babič approach and dynamical ray tracing method in combination with GB summation method [16–18] has found wide applications both in geophysics and in other branches of physics [12,15,19–21]. Not without an influence of Babič approach, the problem of GB diffraction was reduced to the solution of ordinary differential equations of Riccati type also in the framework of *paraxial WKB* approximation, developed in [22,23]. This method became known in plasma physics also as *beam tracing method* [24]. Paraxial WKB stems from representation of the wave field in the geometrical optics form with an additional Gaussian factor.

Eikonal-based paraxial CGO uses the quadratic representation for the eikonal, similar to Babič approach. Quadratic approximation for the complex eikonal, stimulated by Babič solution of the abridged parabolic wave equation, appeared in several publications, dealing with CGO in explicit or inexplicit form [25–31], to name a few. All above-mentioned methods differ in heuristical principles and in techniques of derivation of the Riccati-type equation. Comparative analysis of the Babič approach, paraxial WKB, paraxial CGO, and dynamical ray tracing is presented in paper [32].

The present paper derives the basic relations of the paraxial CGO for 3D Gaussian beams by directly substituting the quadratic form for the complex eikonal into the eikonal equation, generalizing the technique applied for derivation of 2D Riccati equation in [30,31]. We present also two analytically solvable examples of the GB diffraction in 3D space (for homogeneous and for lens-like media) and a non-trivial example of 3D Gaussian beam propagating in an axially symmetrical waveguide medium along a helical ray. The paper is organized as follows. The problem is formulated in Section 2. The eikonal equation in curvilinear ray-centered coordinates, which performs the parallel transport along the central ray, is derived in Section 3 along with the Riccati-type equations for GB complex parameters. Corresponding transport equation for GB amplitude is deduced and solved in paraxial approximation in Section 4. In Section 5 available analytical solutions of the Riccati-type equations are presented for 3D case. Section 6 contains numerical solutions describing a non-trivial problem of GB propagation and diffraction along the helical ray in a circular waveguide. Finally, Section 7 summarizes advantages of the paraxial eikonal-based CGO method.

## 2. Formulation of the problem

Let us consider the propagation of a monochromatic Gaussian beam in a smoothly inhomogeneous isotropic medium characterized by dielectric permittivity $\varepsilon(\mathbf{r})$ in a scalar approximation. The smooth change of the medium parameters implies smallness of the geometrical-optics parameter:

$$\mu_{\text{GO}} \equiv \frac{\lambda}{L} \ll 1. \tag{1}$$

Here $\lambda = 2\pi k_0^{-1} \equiv 2\pi c/\omega$ is the wavelength in vacuum, $\omega$ is the angular frequency, and $L \sim |\nabla \varepsilon|^{-1}$ is the characteristic scale of medium inhomogeneity. Description of a narrow beam within the framework of geometrical optics also implies smallness of the following two parameters:

$$\mu_{\text{DIF}} \equiv \frac{\lambda}{w} \ll 1, \qquad (2)$$

$$\mu_{\text{REF}} \equiv \frac{w}{L} \ll 1, \qquad (3)$$

where $w$ is the characteristic beam width. The diffraction parameter (2) determines the angle of the beam diffraction widening, whereas the refraction parameter (3) characterizes the influence of the medium inhomogeneity on the diffraction. Parameters (2) and (3) change as the beam propagates owing to the variations of the beam's width $w$ (we assume the other quantities to be invariable in order of magnitude). It follows from Eqs. (1)–(3) that

$$\mu_{\text{GO}} \ll \mu_{\text{DIF}}, \mu_{\text{REF}}. \qquad (4)$$

Smallness of three parameters (1)–(3) allows one to use a paraxial approximation along the central ray described by the real geometrical optics. In paraxial approximation the wave field $u(\mathbf{r})$ can be presented in the form as follows

$$u(\tau, \xi_1, \xi_2) = A(\tau) \exp\left[ik_0 \psi(\tau, \xi_1, \xi_2)\right], \quad \psi(\tau, \xi_1, \xi_2) = \psi_c(\tau) + \delta\psi(\tau, \xi_1, \xi_2), \qquad (5)$$

where $\tau$ is the parameter along the central ray, which relates to the ray arc length $\sigma$ as $d\tau = d\sigma / \sqrt{\varepsilon_c}$ and $\xi_{1,2}$ are coordinates orthogonal to the ray, which will be introduced properly in the next Section. Here $\varepsilon_c(\tau) = \varepsilon(\mathbf{r}_c)$ and $\mathbf{r}_c = (\tau, 0, 0)$ is the radius-vector for the central ray in $(\tau, \xi_1, \xi_2)$ coordinates. Amplitude $A(\tau)$ in Eq. (5) determines variations of the beam intensity along the ray. Eikonal $\psi$ in Eq. (5) is supposed to consist of two summands: $\psi_c(\tau)$ is the eikonal on the central ray, while $\delta\psi(\tau, \xi_1, \xi_2)$ is a small deviation from $\psi_c$: $\delta\psi(\mathbf{r}_c) \equiv 0$. Complex eikonal $\delta\psi$ describes both the curvature of the *phase front* of the beam and its *amplitude* profile.

Within paraxial approximation, deviation $\delta\psi$ for GB can be presented as a quadratic form analogously to Babič approach [11–13]:

$$\delta\psi(\tau, \xi_1, \xi_2) = \frac{1}{2} B_{ij}(\tau) \xi_i \xi_j. \qquad (6)$$

In what follows $i = 1, 2$ and summation over repeated indices is implied; $B_{ij}(\tau)$ are complex-valued functions, which constitute a symmetric tensor with $B_{12} \equiv B_{21}$. The real parts of these functions characterize the curvatures of the GB phase front, whereas the imaginary parts determine the elliptical cross-section of the GB (see Eqs. (19) and (20) below). In view of extremal properties of the central ray, the terms linear in $\xi_i$ do not contribute to Eq. (6) in an isotropic medium.

Total complex eikonal $\psi$ satisfies the eikonal equation

$$(\nabla \psi)^2 = \varepsilon, \qquad (7)$$

while amplitude $A$ obeys the transport equation

$$\text{div}\left[A^2 \nabla \psi\right] = 0, \qquad (8)$$

which provides conservation of energy flux along the central ray of the beam. Equations (5)–(8) correspond to the first approximation in geometrical optics parameter (1) and contain a description of the diffraction processes in the first non-vanishing approximation in parameters (2) and (3) (i.e. in the paraxial approximation).

## 3. Complex eikonal equation

It is known that the coordinate system with unit vectors of the Frenet natural trihedron of the central ray, $(\mathbf{l},\mathbf{n},\mathbf{b})$ (here $\mathbf{l}$, $\mathbf{n}$, and $\mathbf{b}$ are the unit vectors of the ray's tangent, principal normal, and binormal respectively), experiences local rotation with respect to the central ray with angular velocity proportional to the ray's torsion $\chi$. As it follows from the Frenet–Serret equations, the coordinate system associated with the natural trihedron is non-orthogonal, so that the corresponding matrix of the Lame coefficients happens to be non-diagonal [33]. Hence it is worth dealing with an orthogonal coordinate system, which would be locally rotationless with respect to the central ray. The unit vectors $(\mathbf{l},\mathbf{e}_1,\mathbf{e}_2)$ of such system are determined by relations [11–13,34]

$$\mathbf{e}_1 = \mathbf{n}\cos\theta + \mathbf{b}\sin\theta, \quad \mathbf{e}_2 = \mathbf{b}\cos\theta - \mathbf{n}\sin\theta, \tag{9}$$

where $\theta$ is the rotation angle, obeying the equation $d\theta/d\tau = -\sqrt{\varepsilon}\chi$. Unit vectors $\mathbf{e}_i$ are transferred along the central ray in accordance with the Levi-Civita parallel transport law (an example of this law in the geometrical optics context is the Rytov-Vladimirskii-Berry law for the evolution of polarization of an electromagnetic wave along the ray, see [35,36]). In what follows we refer to the set $(\mathbf{l},\mathbf{e}_1,\mathbf{e}_2)$ as the Levi-Civita (or parallel transport) basis. If $\boldsymbol{\xi} = \xi_1\mathbf{e}_1 + \xi_2\mathbf{e}_2$ is a two-dimensional vector in the plane orthogonal to the central ray, $\mathbf{r}_c(\tau)$, then the radius-vector $\mathbf{r}$ is connected to the Levi-Civita coordinates $(\tau,\xi_1,\xi_2)$ by relation $\mathbf{r} = \mathbf{r}_c(\tau) + \xi_1\mathbf{e}_1 + \xi_2\mathbf{e}_2$. The Lame coefficients for the Levi-Civita coordinate system $(\tau,\xi_1,\xi_2)$ equal [11–13,34]

$$h_\tau \equiv h = \sqrt{\varepsilon}\left[1 - \frac{(\boldsymbol{\xi}\nabla_\perp)\varepsilon}{2\varepsilon}\right]_{\mathbf{r}=\mathbf{r}_c}, \quad h_{\xi_1} = h_{\xi_2} = 1, \tag{10}$$

where $\nabla_\perp = \partial/\partial\boldsymbol{\xi}$.

By virtue of Eqs. (6) and (10), the eikonal equation (7) in the Levi-Civita coordinates takes the following form:

$$\frac{1}{h^2}\left\{(\dot{\psi}_c)^2 + \dot{\psi}_c\dot{B}_{ij}\xi_i\xi_j + \left(\frac{1}{2}\dot{B}_{ij}\xi_i\xi_j\right)^2\right\} + (B_{1i}\xi_i)^2 + (B_{2i}\xi_i)^2 = \varepsilon. \tag{11}$$

From here on dot stands for differentiation with respect to $\tau$. One can expand dielectric permittivity $\varepsilon$ in equation (11) in a Taylor series in small deviation $\boldsymbol{\xi}$

$$\varepsilon(\mathbf{r}) = \varepsilon_c + (\boldsymbol{\xi}\nabla_\perp)\varepsilon\Big|_{\mathbf{r}=\mathbf{r}_c} + \frac{1}{2}(\boldsymbol{\xi}\nabla_\perp)^2\varepsilon\Big|_{\mathbf{r}=\mathbf{r}_c} + O(\mu_{REF}^3). \tag{12}$$

Within the paraxial approximation all the terms higher than the second-order in small parameters (2) and (3) should be omitted. In this way we neglect in Eq. (11) the third term in the braces, which is proportional to $\xi^4$ and derivatives $\dot{B}_{ij}^2$ and is of order of $\mu_{DIF}^4$. In fact, this approximation corresponds to the parabolic equation in the quasi-optics theory [34].

By solving Eq. (11) with the help of perturbation theory in small parameters (2) and (3) and taking Eq. (12) into account, we have in zero approximation

$$\dot{\psi}_c = \varepsilon_c, \tag{13}$$

Eq. (13) corresponds to the eikonal equation of the real geometrical optics. The Hamiltonian approach determines the central ray's trajectory by the following equations [1,2]:

$$\frac{d(\mathbf{l}\sqrt{\varepsilon})}{d\tau} = \frac{1}{2}\nabla\varepsilon, \quad \frac{d\mathbf{r}}{d\tau} = \mathbf{l}\sqrt{\varepsilon}. \tag{14}$$

In view of extremal properties of the central ray, the terms linear in $\xi_i$ vanish and we obtain a differential equation for the quadratic terms of order of $\mu_{\text{REF}}^2$ and $\mu_{\text{F}}^2$:

$$\dot{B}_{ij}\xi_i\xi_j + (B_{1i}\xi_i)^2 + (B_{2i}\xi_i)^2 = \left[\frac{(\xi\nabla_\perp)^2 \varepsilon}{2} - \frac{3(\xi\nabla_\perp \varepsilon)^2}{4\varepsilon}\right]_{r=r_c}. \quad (15)$$

Equation (15) is a quadratic form in $\xi_i$. In order to satisfy this equation, one should equal all the coefficients in its left-hand and right-hand sides. As a result we obtain a tensor Riccati-type equation for $B_{ij}(\tau)$:

$$\dot{B}_{ij} + B_{ik}B_{kj} = \alpha_{ij}, \quad (16)$$

which is a system of three equations for the complex parameters $B_{ij}(\tau)$:

$$\dot{B}_{11} + (B_{11}^2 + B_{12}^2) = \alpha_{11}, \quad (16a)$$
$$\dot{B}_{12} + B_{12}(B_{11} + B_{22}) = \alpha_{12}, \quad (16b)$$
$$\dot{B}_{22} + (B_{22}^2 + B_{12}^2) = \alpha_{22}. \quad (16c)$$

Here, by analogy with [34] (but with the opposite signs), we have introduced the following quantities:

$$\alpha_{ij}(\tau) = \left(\frac{1}{2}\frac{\partial^2 \varepsilon}{\partial \xi_i \partial \xi_j} - \frac{3}{4\varepsilon}\frac{\partial \varepsilon}{\partial \xi_i}\frac{\partial \varepsilon}{\partial \xi_j}\right)_{r=r_c}. \quad (17)$$

The terms quadratic in $B_{ij}$ in Eqs. (16) are responsible for diffraction in homogeneous medium, while the right-hand side terms of Eqs. (16) describe the influence of the medium inhomogeneity on the diffraction.

Equations (16) are the basic equations for description of GB diffraction in a smoothly inhomogeneous medium. These are ordinary differential equations, which are very useful for the analysis and numerical simulations. This gives a great advantage to the method under consideration over full wave and parabolic wave (quasioptics) equations. Taking into account that refractive index *n* is connected with the phase velocity *v* by a relation $n = c/v$, one can see the equivalence of the Riccati equations (16) and those stemming from abridged parabolic wave equation in the framework of Babič approach [11–13] as well as from dynamical ray tracing method [15] (the later deals with 'associated' ray equations, describing evolution of the second derivatives of the eikonal along the ray).

The complex equations (16) can be presented as a system of six real equations for quantities $R_{ij} \equiv \text{Re}\, B_{ij}$ and $I_{ij} \equiv \text{Im}\, B_{ij}$:

$$\begin{aligned}\dot{R}_{ij} + R_{ik}R_{kj} - I_{ik}I_{kj} &= \alpha_{ij}, \\ \dot{I}_{ij} + R_{ik}I_{kj} + I_{ik}R_{kj} &= 0,\end{aligned} \quad (18)$$

or, in the components:

$$\dot{R}_{11} + R_{11}^2 + R_{12}^2 - I_{11}^2 - I_{12}^2 = \alpha_{11}, \quad (18a)$$
$$\dot{R}_{12} + R_{12}(R_{11} + R_{22}) - I_{12}(I_{11} + I_{22}) = \alpha_{12}, \quad (18b)$$
$$\dot{R}_{22} + R_{22}^2 + R_{12}^2 - I_{22}^2 - I_{12}^2 = \alpha_{22}, \quad (18c)$$
$$\dot{I}_{11} + 2R_{11}I_{11} + 2R_{12}I_{12} = 0, \quad (18d)$$
$$\dot{I}_{12} + R_{12}(I_{11} + I_{22}) + I_{12}(R_{11} + R_{22}) = 0, \quad (18e)$$
$$\dot{I}_{22} + 2R_{22}I_{22} + 2R_{12}I_{12} = 0. \quad (18f)$$

Let us discuss how the above-introduced parameters relate to the real physical quantities. The real and imaginary parts of the quadratic form (6) describe the shapes of the phase front and the beam cross-section respectively. In the paraxial approximation under consideration the phase front in the vicinity of a point $\tau = \tau'$, $\xi_1 = \xi_2 = 0$ represents a second-order surface (a paraboloid or a saddle-type surface in 3D space $(\xi_1, \xi_2, \tau)$) determined by equation $\varepsilon_c(\tau')(\tau - \tau') + \frac{1}{2} R_{ij}(\tau') \xi_i \xi_j = 0$, while the beam cross-section is the second-order curve (ellipse in $(\xi_1, \xi_2)$ plane) determined by equation $k_0 I_{ij}(\tau') \xi_i \xi_j = \text{const}$. Note that in the generic case the principal axes of the real and imaginary parts of (6) do not coincide with each other [39]. If we denote the eigenvalues of tensors $R_{ij}(\tau)$ and $I_{ij}(\tau)$ forms as $R_i(\tau)$ and $I_i(\tau)$, respectively, then in accordance with [37,38] they can be associated with the principal curvature radii of the wave phase front, $\rho_i(\tau)$, and with the beam principal widths (i.e. principal semi-axes of the cross-section ellipse), $w_i(\tau)$, in the following way:

$$R_i = \frac{\sqrt{\varepsilon_c}}{\rho_i}, \quad I_i = \frac{1}{k_0 w_i^2}. \tag{19}$$

In terms of components of $R_{ij}$ and $I_{ij}$ in every point these quantities can be expressed as

$$w_{1,2}^2 = \frac{2}{k_0 \left[ I_{11} + I_{22} \pm \sqrt{(I_{11} - I_{22})^2 + 4 I_{12}^2} \right]}, \tag{20a}$$

$$\rho_{1,2} = \frac{2\sqrt{\varepsilon_c}}{R_{11} + R_{22} \pm \sqrt{(R_{11} - R_{22})^2 + 4 R_{12}^2}}. \tag{20b}$$

Equation (20a) introduces an additional restriction for CGO description of GB. Indeed, if one requires $w_i^2$ to be positive, this leads to inequality

$$I_{11} + I_{22} > \sqrt{(I_{11} - I_{22})^2 + 4 I_{12}^2}, \tag{21}$$

which ensures that $I_i > 0$ is positive, i.e. the beam is bounded.

## 4. Transport equation for amplitude and conservation of energy flux

In the ray-centered coordinates $(\tau, \xi_1, \xi_2)$ the transport equation (8) takes the form

$$\frac{2}{h^2} \frac{\partial \psi}{\partial \tau} \frac{dA}{d\tau} + \left[ \frac{1}{h} \frac{\partial}{\partial \tau} \left( \frac{1}{h} \frac{\partial \psi}{\partial \tau} \right) + \frac{\partial^2 \psi}{\partial \xi_1^2} + \frac{\partial^2 \psi}{\partial \xi_2^2} \right] A = 0. \tag{22}$$

By using paraxial approximation and introducing new amplitude $\tilde{A} = \varepsilon_c^{1/4} A$, the above equation can be reduced to the following form

$$\dot{\tilde{A}} + \frac{1}{2} \operatorname{Tr} B_{ij} \tilde{A} = 0, \tag{23}$$

where $\operatorname{Tr} B_{ij} \equiv B_{ii} = B_{11} + B_{22}$. It admits an explicit solution

$$\tilde{A}(\tau) = \tilde{A}_0 \exp\left[ -\frac{1}{2} \int \operatorname{Tr} B_{ij} \, d\tau \right], \tag{24}$$

where $\tilde{A}_0 \equiv \tilde{A}(0)$ is the initial amplitude of the beam. Thus, tensor $B_{ij}$, defined from Eqs. (16), enables one to determine the complex amplitude $A$ as well.

The absolute value of $\tilde{A}$ equals

$$\left|\tilde{A}\right|=\left|\tilde{A}_0\right|\exp\left[-\frac{1}{2}\int\mathrm{Tr}\,R_{ij}\,d\tau\right]. \qquad (25)$$

It follows from Eqs. (18d)–(18f) that the combination $D \equiv \mathrm{Det}(I_{ij}) = I_{11}I_{22} - I_{12}^2$ obeys the equation

$$\dot{D} + 2\,\mathrm{Tr}\,R_{ij}\,D = 0. \qquad (26)$$

Therefore

$$\int\mathrm{Tr}\,R_{ij}\,d\tau = -\ln\sqrt{\frac{D}{D_0}}, \qquad (27)$$

where $D_0 \equiv D(0)$. Then, Eqs. (25) and (27) yield

$$\left|\tilde{A}\right|^2 = \left|\tilde{A}_0\right|^2\sqrt{\frac{D}{D_0}}. \qquad (28)$$

Note, that $1/\sqrt{D}$ has the sense of the area of the GB cross-section: $1/\sqrt{D} \propto w_1 w_2$, so that Eq.(28) takes the form

$$|\tilde{A}|^2 = |\tilde{A}_0|^2\,\frac{w_{10}w_{20}}{w_1 w_2}. \qquad (28a)$$

The above equations express the conservation of the energy flux through the GB cross-section. The total energy flux $\Pi$ in the beam can be calculated by integration of Eq. (8) together with (5) and (6):

$$\Pi \equiv \iint \sqrt{\varepsilon}\,|u|^2\,d\xi_1 d\xi_2 \propto \frac{|\tilde{A}|^2}{\sqrt{D}} \propto |\tilde{A}|^2\,w_1 w_2 = \mathrm{const}, \qquad (29)$$

where the integration is performed over the whole plane $(\xi_1,\xi_2)$ transverse to the ray. The energy flux conservation similar to Eq. (28) has been obtained analytically also in [34] within the frame of the parabolic equation approach for the case of separated variables, when $\alpha_{12} \equiv 0$, which corresponds to the torsionless ray. These results are in agreement with the description of GB diffraction in the framework of abridges parabolic wave equation [11–13,34] as well as with dynamical ray tracing method [14–16] and paraxial WKB approximation [22–24].

It is worth noticing that in the presence of weak absorption ($\mathrm{Im}\,\varepsilon \ll \mathrm{Re}\,\varepsilon$) the dissipation factor $\exp\left[-(k/2)\int\mathrm{Im}\,\varepsilon\,dl\right]$ appears in Eq. (24), like in the traditional geometrical optics [1,2].

## 5. Analytical solutions

**5.1. *Diffraction in a homogeneous medium*.** In a homogeneous medium, where $\alpha_{ij} \equiv 0$, equations (16) possess an analytical solution. In the general case it can be written in a matrix form [11,12,27]:

$$\hat{\mathbf{B}}(\tau) = \left[\hat{\mathbf{B}}^{-1}(0) + \hat{\mathbf{I}}\tau\right]^{-1}, \qquad (30)$$

where $\hat{\mathbf{B}}$ is $2\times 2$ matrix with components $B_{ij}$, while $\hat{\mathbf{I}}$ is the unit matrix. In particular, when the principal axes of the phase front and of the beam cross-section (i.e. those of tensors $R_{ij}$ and $I_{ij}$) do not coincide, solution (30) describes evolution of GB with general astigmatism [39].

Let us consider the special case when the principal axes of $R_{ij}$ and $I_{ij}$ coincide with each other at $\tau = 0$. Given this, one can choose axes of $\xi_i$ coordinates in such a way that $B_{12}(0) = 0$. Then it follows from Eq. (16b) that $B_{12}(\tau) \equiv 0$, $B_{11}(\tau) \equiv B_1(\tau)$, and $B_{22}(\tau) \equiv B_2(\tau)$, i.e. the cross-section and the wave front of the beam under consideration conserve their orientation along the whole trajectory. Then solution (30) yields:

$$B_i = \frac{B_{i0}}{1 + B_{i0}\tau}, \quad (31)$$

where $B_{i0} \equiv B_i(0)$. By substituting Eq. (31) in Eq. (23), we obtain the equation for amplitude $\tilde{A}$:

$$\dot{\tilde{A}} + \frac{1}{2}\left(\frac{B_{10}}{1 + B_{10}\tau} + \frac{B_{20}}{1 + B_{20}\tau}\right)\tilde{A} = 0, \quad (32)$$

which yields

$$\tilde{A} = \frac{\tilde{A}_0}{\sqrt{(1 + B_{10}\tau)(1 + B_{20}\tau)}} = \tilde{A}_0 \sqrt{\frac{B_1 B_2}{B_{10} B_{20}}}. \quad (33)$$

Let us express solutions (31) and (33) in terms of Eq. (19). Assuming for the simplicity sake the initial phase front of the beam to be plane, $\rho_{i0} \equiv \rho_i(0) = \infty$, and substituting Eq. (19) with $B_i = R_i + iI_i$ into Eq. (31), we come to the known expressions for the evolution of $\rho_i$ and $w_i$:

$$\rho_i = \tau\left[1 + \left(\frac{L_{Ri}}{\tau}\right)^2\right], \quad w_i = w_{i0}\sqrt{1 + \left(\frac{\tau}{L_{Ri}}\right)^2}, \quad (34)$$

where $L_{Ri} = k_0 w_{i0}^2$ is the Rayleigh distance, corresponding to the initial width $w_{i0} \equiv w_i(0)$.

Equations (34) demonstrate typical behavior for GB diffraction in empty space. In the near zone, $\tau \ll L_{Ri}$, GB widths are approximately constant, whereas the curvature radii of the wave front decrease:

$$w_i \approx w_{i0} = \text{const}, \quad \rho_i \approx \frac{L_{Ri}^2}{\tau}. \quad (35a)$$

In the far zone, $\tau \gg L_{Ri}$, both the beam widths and the curvature radii of the wave front increase linearly:

$$w_i \approx \frac{w_{i0}}{L_{Ri}}\tau, \quad \rho_i \approx \tau. \quad (35b)$$

At the Rayleigh distance $\tau = L_{Ri}$, the $i$th curvature radius reaches its minimum value, $\rho_{i\min} = 2L_{Ri}$, with the corresponding beam's width equaling $\sqrt{2}w_{i0}$.

By substituting representation (19) into Eq. (33), we find that in the near zone (with the plane initial phase front) the absolute value of complex amplitude $A$ remains almost constant:

$$|A| \approx |A_0| = \text{const}, \quad (36a)$$

while in the far zone

$$|A| \approx \frac{k_0 w_{10} w_{20}}{\tau}|A_0| = \sqrt{\frac{w_{10} w_{20}}{w_1 w_2}}|A_0|. \quad (36b)$$

Thus, Eqs. (36) agree with the conservation of the energy flux through the beam cross-section, Eqs. (28) and (29).

*5.2. Diffraction in a lens-like medium.* Let us turn to the analysis of equations (16) in an inhomogeneous medium. A detailed analysis of the influence tensor $\alpha_{ij}$ on the beam dynamics within the framework of a parabolic equation was presented in paper [34]. When the tensor $\alpha_{ij}$ has negative eigenvalues, $\alpha_i < 0$, then the inhomogeneous medium focuses the beam, thereby compensating to some degree the diffraction widening. On the contrary, if $\alpha_{ij}$ possesses positive eigenvalues, $\alpha_i > 0$, then the inhomogeneous medium defocuses the beam.

GB diffraction in 2D lens-like medium with $\varepsilon = 1 \mp \xi^2 / L^2$ (waveguide and anti-waveguide respectively) was studied in paper [29] for a case, in which the beam propagates along the medium symmetry axis. In this Section we extend the results of paper [29] to 3D lens-like medium and consider a beam propagating along the axis of a waveguide (or an anti-waveguide) with $\varepsilon = 1 + \alpha_1 \xi_1^2 + \alpha_2 \xi_2^2$. We assume that the principal axes of the tensors $\alpha_{ij}$, $R_{ij}$, and $I_{ij}$ coincide with each other. For lens-like media of this kind the right-hand sides of Eqs. (16) become constant: $\alpha_{11}(\tau) \equiv \alpha_1(\tau) = \text{const}$, $\alpha_{22}(\tau) \equiv \alpha_2(\tau) = \text{const}$, and $\alpha_{12}(\tau) = \alpha_{21}(\tau) \equiv 0$. Then Eqs. (16) take the form of two independent equations:

$$\dot{B}_i + B_i^2 = \alpha_i, \tag{37}$$

which are identical to the equation analyzed in [29] for 2D medium. Therefore, all the results of [29] can be applied to 3D medium under consideration. Solution of Eq. (37) can be presented in the following form

$$B_i = \sqrt{\alpha_i} \frac{C_i e^{2\sqrt{\alpha_i}\tau} - 1}{C_i e^{2\sqrt{\alpha_i}\tau} + 1}, \quad \text{where} \quad C_i = \frac{\sqrt{\alpha_i} + B_{i0}}{\sqrt{\alpha_i} - B_{i0}}. \tag{38}$$

which embraces both the waveguide (negative $\alpha_i$) and anti-waveguide (positive $\alpha_i$) propagation. For the plane initial phase front, when $\rho_{i0} = \infty$, one has $B_{i0} = i / k_0 w_{i0}^2 = i / L_{Ri}$.

Let us determine the characteristic inhomogeneity scales along $\xi_i$ axes as $L_i = |\alpha_i|^{-1/2}$. Then, in the case of defocusing medium, $\alpha_i > 0$, at the distances $\tau \geq L_i$ the defocusing processes start to prevail and the beam widens in an exponential way:

$$w_i \approx \frac{w_{i0}}{2} \sqrt{1 + \left(\frac{L_i}{L_{Ri}}\right)^2} e^{\tau/L_i}. \tag{39}$$

At a certain stage condition (3) fails and solutions (38) and (39) become inapplicable. From the analysis of Eq. (39) one can derive restrictions for the distance of CGO and paraxial approximation applicability:

$$\tau \leq L_i \min\left(\ln \mu_{\text{REF}}^{-1}, \ln \mu_{\text{DIF}}^{-1}\right). \tag{40}$$

In practice, the logarithms of small parameters (2) and (3) cannot be too large, therefore in the case of defocusing medium the CGO solutions (as well as those of parabolic equation) are applicable at distances of a few inhomogeneity scales only.

For a focusing lens-like medium, $\alpha_i < 0$, solution (38) takes the form

$$B_i = \frac{1}{L_i} \frac{i \dfrac{L_i}{L_{Ri}} - \tan(\tau/L_i)}{i \dfrac{L_i}{L_{Ri}} \tan(\tau/L_i) + 1}. \tag{41}$$

The corresponding beam widths $w_i$ and the wave front curvature radii $\rho_i$, Eqs. (19), are as follows:

$$w_i = w_{i0} \sqrt{1 + \left[\left(\frac{L_i}{L_{Ri}}\right)^2 - 1\right] \sin^2(\tau/L_i)}, \tag{42a}$$

$$\rho_i = L_i \frac{1 + \left(\dfrac{L_i}{L_{Ri}}\right)^2 \tan^2(\tau/L_i)}{\left[\left(\dfrac{L_i}{L_{Ri}}\right)^2 - 1\right] \tan(\tau/L_i)}. \tag{42b}$$

If $L_i = L_{Ri}$, the balance between diffraction broadening and focusing process takes place. In this case, equation (37) possesses a stationary solution, according to which GB with flat wave front ($\rho_i \to \infty$) propagates with constant width $w_i = w_{i0}$. The stationary solution corresponds to the region where $\mu_{REF} \sim \mu_{DIF}$ and CGO method is applicable for an arbitrary large distance $\tau$. In a general (non-stationary) case the beam parameters oscillate around the equilibrium values according to Eqs. (42). According to [29], these oscillations are caused by the beating between eigen-modes of the waveguide. Oscillating CGO solutions are also applicable for arbitrary large distances.

## 6. Gaussian beam diffraction along helical ray in an axially inhomogeneous medium

In this Section we consider a non-trivial example of the GB propagation and diffraction in an essentially 3D medium with inseparable variables. Non-zero torsion of the central ray causes a non-trivial parallel transport, Eq. (9), and provides for entanglement of equations, because the inhomogeneity influence, i.e. the right-hand sides of Eqs. (16), is linked up to normal-binormal coordinate system ($\nabla \varepsilon$ lies in the $(\mathbf{l}, \mathbf{n})$ plane, see [1,2] and Eqs. (14)), while the diffraction by itself, i.e. the left-hand sides of Eqs. (16), is attached to the parallel transport coordinate system (9). Thus, diffraction of a ray with torsion demonstrates some sort of 'competition' between the normal-binormal symmetry of inhomogeneity and parallel transport symmetry of the pure diffraction.

Let us consider axially symmetrical focusing medium in cylindrical coordinates ($r, \varphi, z$) with dielectric permittivity

$$\varepsilon(r) = \varepsilon_0 - \frac{r^2}{L^2} > 0. \qquad (43)$$

Such a medium forms a 3D circular waveguide, which manifests focusing properties in radial direction, because

$$\alpha_r \equiv \frac{\varepsilon''}{2} - \frac{3\varepsilon'^2}{4\varepsilon} = -\frac{1}{L^2} - \frac{3r^2}{\varepsilon L^4} < 0,$$

where primes stand for the derivatives with respect to $r$. At the same time the medium is homogeneous in azimuthal and longitudinal directions, $\alpha_{\varphi,z} \equiv 0$, and can bring about the unbounded diffraction widening in these directions. In the simplest case, when the beam propagates along the waveguide axis $z$, one deals with a lens-like medium of cylindrical symmetry with $\alpha_{ij} = \text{const} > 0$. The beam's widths will experience periodical compressions and stretches, like in Eq. (42a).

Of much more interest is the case of a beam propagating along the helical ray laying on the cylindrical surface of constant radius $r_c$. In contrast to all the above-considered examples, such a ray demonstrates a non-trivial parallel transport (9). From Eqs. (14) (see also [1,2]) one can find out that radius $r_c$ equals

$$r_c = -l_\varphi^2 \left.\frac{2\varepsilon}{\varepsilon'}\right|_{r=r_c}. \qquad (44)$$

Here $l_\varphi$ is the axial component of the unit tangent vector $\mathbf{l}$. Substitution of Eq. (43) into Eq. (44) yields

$$r_c = \frac{l_\varphi \sqrt{\varepsilon_0}}{\sqrt{1 + l_\varphi^2}} L. \qquad (45)$$

The torsion of this helical ray equals

$$\chi = \frac{l_z l_\varphi}{r_c} = \frac{\sqrt{1-l_\varphi^4}}{\sqrt{\varepsilon_0} L} = \text{const}, \qquad (46)$$

where $l_z = \sqrt{1-l_\varphi^2}$ since $l_r = 0$.

The normal and binormal to the ray are connected to the unit vectors of the cylindrical coordinates as

$$\mathbf{n} \equiv -\mathbf{e}_r, \quad \mathbf{b} = -\left(l_z \mathbf{e}_\varphi - l_\varphi \mathbf{e}_z\right), \qquad (47)$$

By using Eq. (9) we introduce coordinate system corresponding to the Levi-Civita parallel transport:

$$\mathbf{e}_1 = -\mathbf{e}_r \cos\theta - \left(l_z \mathbf{e}_\varphi - l_\varphi \mathbf{e}_z\right)\sin\theta, \quad \mathbf{e}_2 = -\left(l_z \mathbf{e}_\varphi - l_\varphi \mathbf{e}_z\right)\cos\theta + \mathbf{e}_r \sin\theta. \qquad (48)$$

Here

$$\theta = -\frac{\sqrt{1-l_\varphi^2}}{L}\tau \qquad (49)$$

is the rotation angle of the normal $\mathbf{n}$ around the ray.

To solve equations (16) it is necessary to determine functions $\alpha_{ij}(\tau)$ in the coordinate system $(\xi_1, \xi_2, \tau)$ with unit vectors (48). Taking into account that $\partial/\partial\xi_i = (\mathbf{e}_i \nabla)$ and that for an axially inhomogeneous medium $\nabla = \mathbf{e}_r d/dr$, one can rewrite Eq. (17) as

$$\alpha_{ij} = (\mathbf{e}_i \mathbf{e}_r)(\mathbf{e}_j \mathbf{e}_r)\left[\frac{\varepsilon''}{2} - \frac{3(\varepsilon')^2}{4\varepsilon}\right]. \qquad (50)$$

In view of Eqs. (45) and (48) these expressions take the form

$$\alpha_{11} = -\frac{1+3l_\varphi^2}{L^2}\cos^2\theta, \quad \alpha_{22} = -\frac{1+3l_\varphi^2}{L^2}\sin^2\theta, \quad \alpha_{12} = \alpha_{21} = \frac{1+3l_\varphi^2}{L^2}\sin\theta\cos\theta. \qquad (51)$$

We see that the inhomogeneity is linked up to the cylindrical or the normal-binormal coordinate system, Eq. (47), and in the parallel transport coordinates, Eq. (48), the tensor $\alpha_{ij}$, Eq. (51), rotates resulting in the entanglement of equations (16a)–(16c).

Equations (16) with (49), (51) have been solved numerically, and taking Eqs. (20) into account we can present the dynamics of the GB parameters. Figure 1 depicts the evolution of the helical GB's widths $w_i(\tau)$, Eq. (20a). It is seen that one of the widths grows infinitely, whereas another one demonstrates oscillations typical for focusing media. Such a behavior confirms a phenomenon that has been predicted in [34]: when the value of the squared beam's torsion turns out between the eigenvalues of tensor $-\alpha_{ij}$, the beam becomes parametrically unstable. In our case we have $-\alpha_{\varphi,z} < \chi^2 < -\alpha_r$ and unlimited growth of one of the beam's width can be regarded as a consequence of this instability. Due to such instability CGO solutions are applicable only at finite distances of the order of several Rayleigh distances: after that the condition (21) fails. Figure 2 shows the evolution of the curvature radii of the beam's phase front. Both radii oscillate with the period of small width oscillations, Fig. 1a. At that, there are regions of equal signs of the curvatures (in this case phase front is a paraboloid) as well as regions where their signs are opposite (phase front is a saddle-type surface). Figure 3 depicts explicitly the evolution of the beam cross-section (it is determined by equation $k_0 I_{ij}\xi_i\xi_j = 1$) and the cross-section of its phase front (which we determine by $k_0 R_{ij}\xi_i\xi_j = 1$) in the Levi-Civita coordinates $(\xi_1, \xi_2)$ associated with the unit vectors (48), whereas Fig. 4 shows the same evolutions in the coordinates $(\xi_n, \xi_b)$ connected to the normal-binormal unit vectors (47) (in fact, in the coordinates attached to the cylindrical ones). One can see that the beam's cross-section rotates in $(\xi_1, \xi_2)$ coordinates with angular velocity that approximately equals $d\theta/d\tau = -\sqrt{\varepsilon}\chi$, Eqs. (46), (49). At the same time, in the $(\xi_n, \xi_b)$ coordinates the cross-section ellipse approximately conserves its orientation being stretched along $\xi_b$ axis. This result means that the diffraction of the GB under consideration is determined mainly by the inhomogeneity. As was mentioned above, the focusing character of the radial inhomogeneity leads to finite oscillations of the beam's width along $r$ (i.e. $\xi_n$) together with infinite grows of the beam along $\xi_b$ axes, along which the medium is locally homogeneous. However, the beam's cross-section ellipse is not exactly oriented along $\xi_b$ axes: as it is seen at Fig. 3, it is slightly inclined in the positive (clockwise) direction. This reflects a natural tendency of the beam to follow the parallel transport law. Indeed, in $(\xi_n, \xi_.)$-coordinates the parallel transport coordinates $(\xi_., \xi_.)$ rotate in the positive direction, and the inclined orientation of the cross-section ellipse reflects some sort of balance between the dominated inhomogeneity symmetry and the parallel transport tendency. As for phase front evolution, it demonstrates rather complicated dynamics that can be attached neither to parallel transport coordinate system nor to the normal-binormal one. In Fig. (3) and Fig. (4) one can see explicitly the cases of paraboloid phase front (elliptical dashed curves) and saddle-type phase front (hyperbolic dashed curves). All mentioned pictures completely describe the helical GB's complex evolution, thus demonstrating the power of the paraxial CGO approach.

## 7. Conclusion

In summary, we have considered a problem of Gaussian beam (GB) diffraction in a 3D inhomogeneous medium. The main result of the paper consists in the derivation of the paraxial complex geometrical optics relations from the first principles, which reduces the diffraction problem to the solution of ordinary differential equations, similar to Babič approach, dynamical ray tracing and paraxial WKB.

We have considered various examples of the GB diffraction, including ones analytically solvable with separation of variables as well as a non-trivial one. In the homogeneous medium and in the lens-like medium with separable variables our method leads to known equations for GB diffraction and to natural extension of the previously studied 2D case. The main non-trivial example is the diffraction of GB propagating in axially symmetrical waveguide medium along a helical ray, which is of great interest in connection with the Berry phase and parallel transport of the polarization in the electromagnetic vector problem [36]. In contrast to the problem of the polarization evolution and to the example of torsion-less beams in [34], we have found that the evolution of GB cross-section is determined at large distances mainly by the inhomogeneity symmetry, rather than by the parallel transport law. At that, the parallel transport influence is also noticeable and reveals itself in the form of the incline of the cross-section ellipse in the direction of rotation of the parallel transport coordinates with respect to normal-binormal coordinates.

## Acknowledgements

The work was partially supported by INTAS (grant No. 03-55-1921). The authors are indebted to R. Nowack, L. Klimes, and I. Psencik, for friendly support and valuable references on Gaussian beams theory. K.B. is grateful to Maritime University of Szczecin for hospitality.

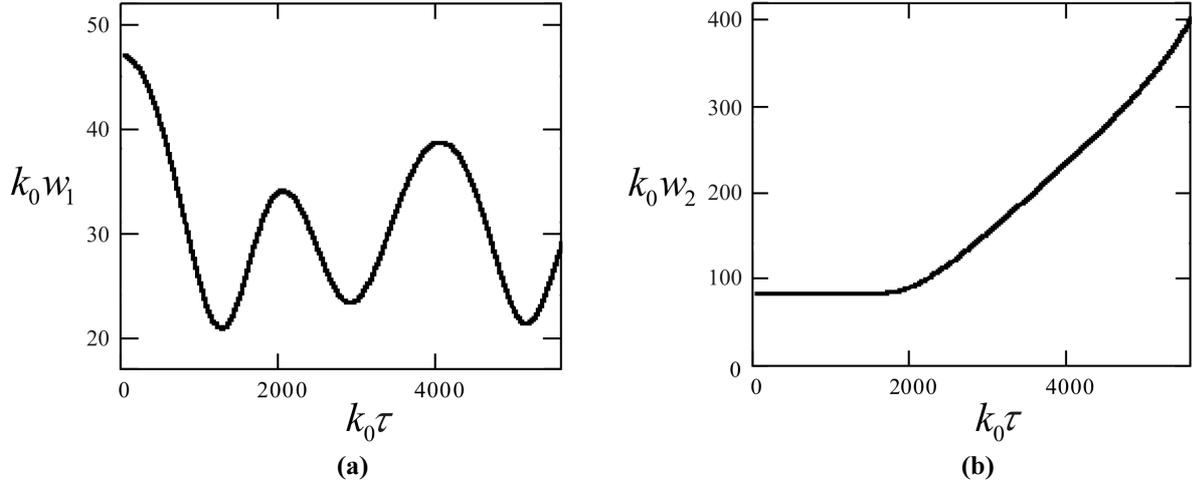

**Fig. 1.** Numerical results for widths $w_{1,2}$ vs. distance $\tau$ (from 0 to $3L_{R1}$, where $L_{R1} = \pi w_{10}^2 / \lambda$ is the Rayleigh distance for smaller width) for the Gaussian beam propagating along the helical ray described in Section 6. The parameters are as follows: $\varepsilon_0 = 1$, $L = 200\lambda$, $l_\varphi = 1/\sqrt{2}$, $w_{10} = 10\lambda$, $w_{20} = 15\lambda$, and the initial phase front is flat: $\rho_{i0} = \infty$.

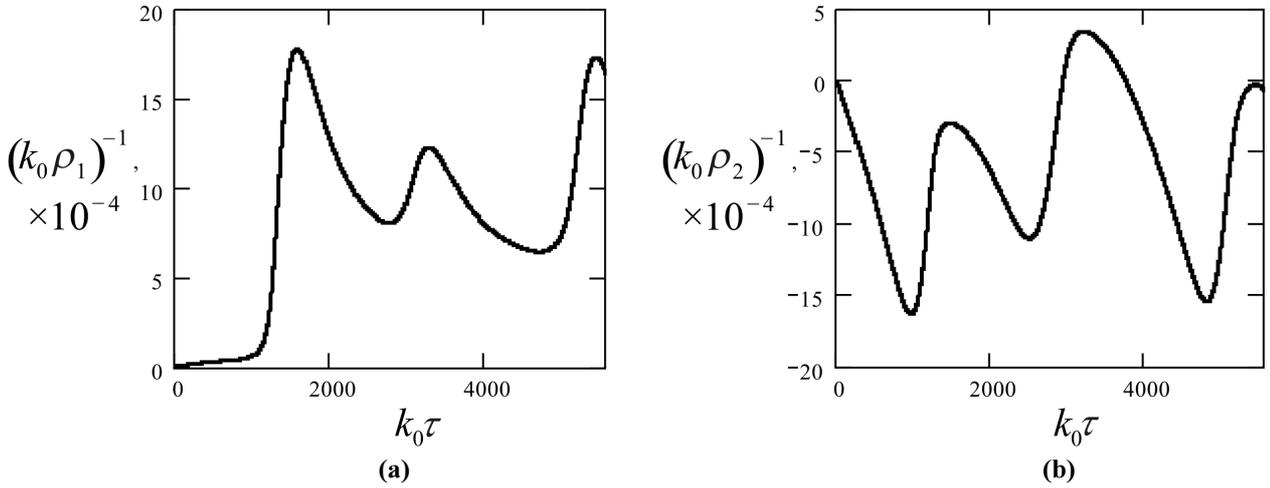

**Fig. 2.** The GB phase front curvatures $\rho_i^{-1}$ vs. distance $\tau$. The interval and parameters are the same as in Fig. 1.

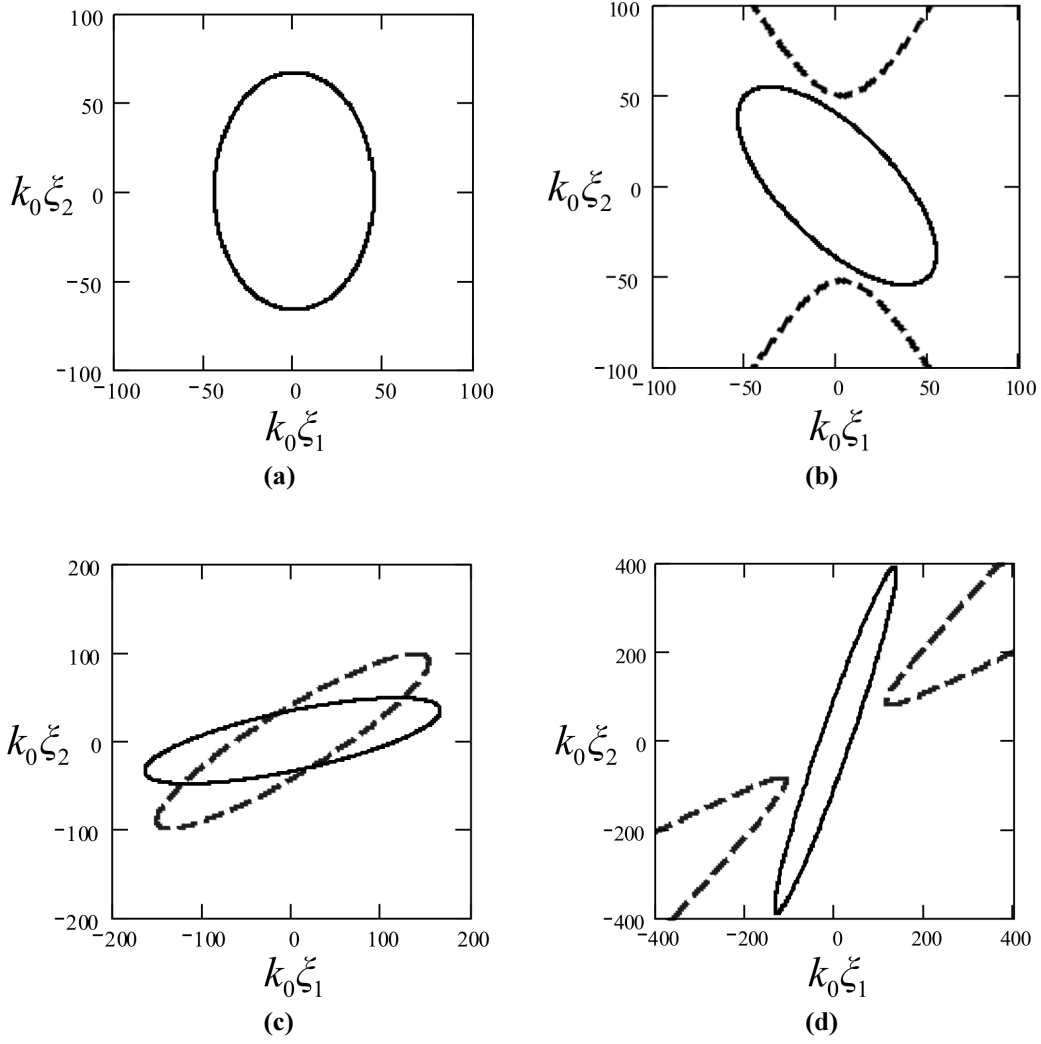

**Fig. 3.** The evolutions of the intensity and the phase front cross-sections (solid and dashed curves, respectively) for the Gaussian beam propagating along the helical ray in the Levi-Civita (parallel transport) coordinates. The parameters are the same as in Fig. 1, pictures (a)–(d) correspond to the values of $\tau$, for which the angle $\theta$, Eq. (48), equals $0$, $-\pi/3$, $-2\pi/3$, $-\pi$, respectively. For the given parameters' values this range of $\tau$ equals to that in Figs. 1 and 2.

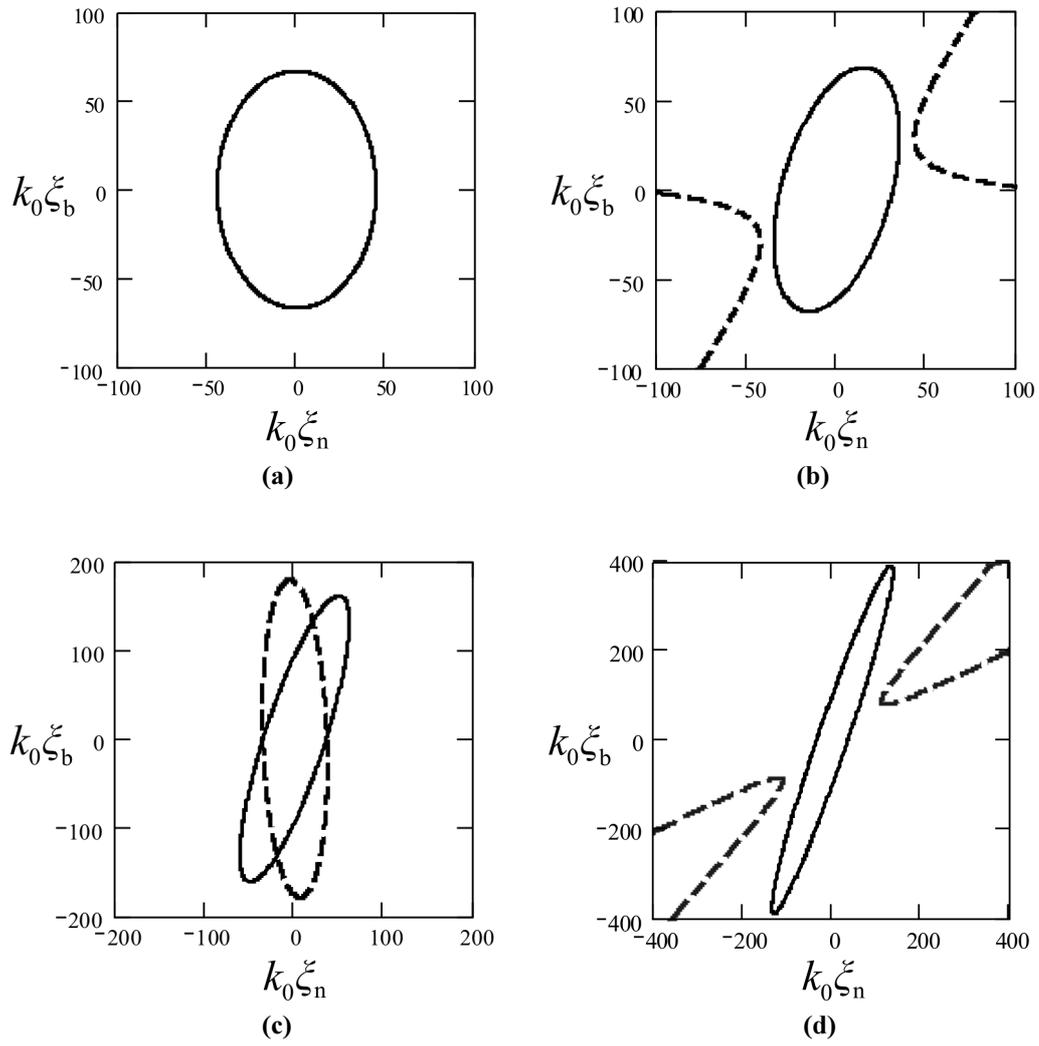

**Fig. 4.** The same as in Fig. 3 but in the coordinates attached to the normal and binormal, Eq. (47).